\documentclass{article}
\usepackage{acra}

\usepackage{siunitx}
\usepackage{graphicx}
\usepackage{xcolor}
\usepackage{float}
\usepackage{amsmath}
\usepackage{amssymb}
\usepackage{cleveref}
\usepackage{booktabs}
\usepackage{algorithm}
\usepackage{algpseudocode}
\RequirePackage[format=plain,labelformat=simple,labelsep=period,font=small,compatibility=false]{caption}
\RequirePackage[font=footnotesize,skip=3pt,subrefformat=parens]{subcaption}
\title{Beyond Visibility: Real-Time Surface Accessibility Fields from Sparse LiDAR}
\author{Bradley Scott, Sam Schofield, Richard Green \\ Computer Science and Software Engineering, University of Canterbury, New Zealand \\ 
bradleyg.scott@pg.canterbury.ac.nz}

\begin{document}

\maketitle

\begin{abstract}
Understanding which surfaces a given tool can physically access is fundamental to robotic interaction, yet 3D perception typically stops at geometric reconstruction or visibility estimation. Visibility does not imply accessibility: a visible point may still be blocked by tool geometry or approach-corridor clearance. We propose the Accessibility Field: a per-point surface accessibility labelling, computed from streaming sparse LiDAR and updated at sensor rate as the platform moves. Our GPU-based method evaluates surface points against precomputed tool-geometry kernels across rotated approach orientations, underpinned by a scan-centric TSDF integration scheme suited to non-repetitive sensors such as the Livox Mid-360. The system is tool-agnostic, requires no prior scene model, and runs in real time on both workstation and Jetson Orin edge hardware. On synthetic primitives, our method achieves F1\,=\,90.8\,\%, against 69.8\,\% for a Hidden Point Removal baseline; ablations within our own pipeline give 81.5\,\% replacing the TSDF query with a point-cloud collision check, and 33.9\,\% replacing scan-centric integration with projective, nvblox-style integration. On a mature synthetic \textit{Pinus radiata} model, the method labels 56.8\,\% of visible pine branch surfaces as physically inaccessible.
\end{abstract}    
\section{Introduction}
\label{sec:intro}

Three-dimensional scene understanding has advanced rapidly, from raw geometry reconstruction towards richer scene representations, including occupancy information, semantics, and instance identification. Yet for systems that must physically interact with the world, a perceptual gap remains: we should not only know where surfaces are, but also understand their capacity for interaction with a given tool. In our work, we call this \textbf{surface accessibility evaluation}: a step beyond geometric reconstruction that can enable robotic systems to make explicit, precise decisions about where and how a given tool may act within a scene.

\begin{figure}[t]
  \centering
    \includegraphics[width=\linewidth]{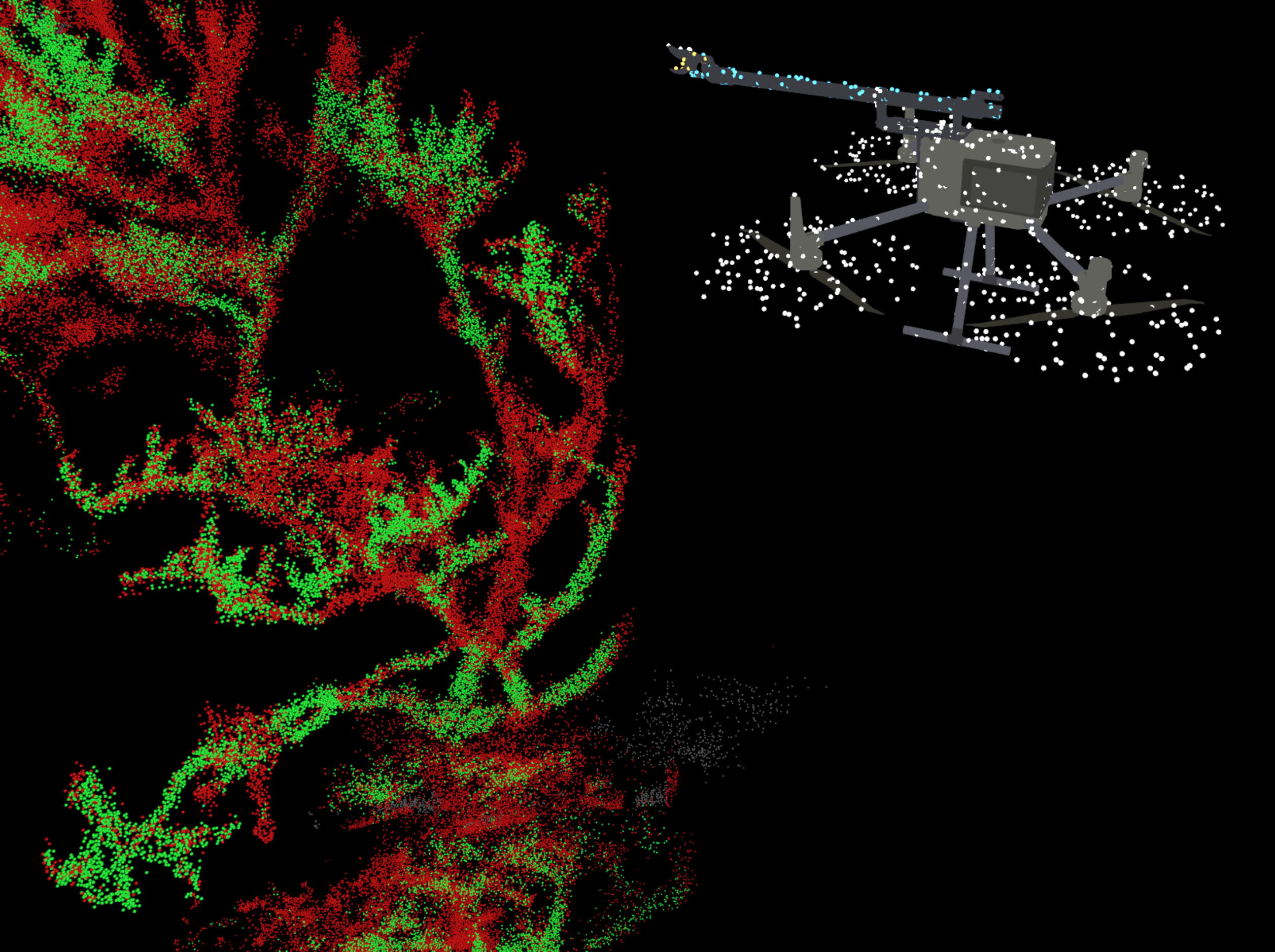}
    \caption{\textbf{Accessibility Field.} Per-point surface accessibility computed from a Livox Mid-360 LiDAR scan of an outdoor scene. Colours indicate point classification status. \textit{Green}: points accessible to the UAV pruning tool; \textit{red}: points blocked by surrounding geometry along the approach corridor; \textit{grey}: below the confidence threshold ($c < 0.2$) and not yet scored, typically distant or occluded canopy with insufficient LiDAR returns.}
    \label{fig:access_main}
\end{figure}

Environment accessibility is not a novel concept: prior research has investigated precomputed workspace measures for fixed-base robotic arms~\cite{zachariasCapturingRobotWorkspace2007,vahrenkampManipulabilityAnalysis2012,vahrenkampRepresentingRobotsWorkspace2015}, and the term \textit{Reachable Workspace} has been applied as an ergonomic outcome measure in patient rehabilitation~\cite{hanValidityReliabilitySensitivity2013,leeUpperLimbThreeDimensional2020}. Both approaches, however, assume a known environment and a fixed kinematic root node. For mobile platforms such as Unmanned Aerial Vehicles (UAVs) operating in complex, unstructured outdoor scenes, neither assumption holds. Localisation and mapping must be performed incrementally from live sensor data~\cite{zhangLOAMLidarOdometry2014,shanLIOSAMTightlycoupledLidar2020}, and the accessibility field must update continuously as new surfaces are observed.

Existing approaches to accessibility are either object-centric and learned for a limited range of interactions, or limited to visibility, determining whether a point can be seen from a given viewpoint. Visibility does not imply accessibility: a point visible from multiple viewpoints may still be blocked by tool geometry along the approach path, or lie in a region with limited observability. In this paper, we make this distinction and address it directly.

We present a real-time system that produces a continuously updated accessibility field over a live LiDAR point cloud, classifying each point by its physical accessibility for a given tool as the robot moves through the scene (\Cref{fig:access_main}).\\

\noindent~Our contributions are as follows:

\begin{itemize}
    \item An \textit{accessibility field}: a novel 3D scene understanding output providing per-point surface accessibility from sparse LiDAR, updated incrementally at sensor rate.
    \item A GPU-accelerated, scan-centric TSDF integration approach with positive-clamped fusion, robust to sparse non-repeating scan patterns and multi-view thin structure ambiguity.
    \item A synthetic evaluation of accessibility labels across simple objects and real-world scale tree models, with qualitative validation on real Livox Mid-360 environment scans and runtime and memory profiling confirming real-time operation on workstation and NVIDIA Jetson Orin edge hardware.
\end{itemize}

\section{Related work}
\label{sec:background}

Prior work has addressed the key problems underlying surface accessibility in isolation: geometric accessibility analysis, volumetric environment representation, and point visibility, yet no existing method combines them into a scene-level accessibility field.

\subsection{Geometric Accessibility Analysis}
\label{subsec:geoaccess}

Geometric accessibility analysis uses visibility maps~\cite{elberAccessibility5axisMilling1994,wooVisibilityMapsSpherical1994,houComputingGlobalVisibility2017} and C-space accessibility
cones~\cite{spyridiAccessibilityAnalysisAutomatic1990,limaiemGeneralMethodAccessibility1997} to determine collision-free tool interactions for subtractive manufacturing (CNC machining). GPU-accelerated sampling~\cite{balasubramaniamGenerating5axisNC2000} and KD-tree spatial partitioning~\cite{alvarezAccessibilityAnalysisAutomatic2008} reduce intersection complexity, though these methods remain susceptible to shallow-pocket errors on high-curvature geometries. Critically, all such methods require a complete, noise-free surface mesh, which is incompatible with incrementally observed, noisy point clouds from live LiDAR sensors. Classical reachability and trajectory-planning simulators face the same barrier from the other direction: they compute collision-free motions via forward kinematics over a known, complete environment
model~\cite{zachariasCapturingRobotWorkspace2007,vahrenkampManipulabilityAnalysis2012,vahrenkampRepresentingRobotsWorkspace2015}, assuming a fixed kinematic root and a static, fully observed scene. Neither assumption holds for a mobile platform mapping incrementally from live sensor data. Discretised volumetric representations avoid these requirements, enabling
incremental updates from streaming sensor data.

\subsection{Volumetric Environment Representation}
\label{subsec:vol_env_rep}
Volumetric distance fields provide the spatial foundation required for collision-aware accessibility evaluation. Early Truncated Signed Distance Fields (TSDFs), such as KinectFusion~\cite{newcombeKinectFusionRealtimeDense2011}, enable dense environment reconstruction but lack global distance information outside the truncation band and incur high memory overheads at scale. Voxblox~\cite{oleynikovaVoxbloxIncremental3D2017} addressed both limitations with a hierarchical hashed voxel structure for incremental Euclidean Signed Distance Field (ESDF) generation, though its TSDF-to-ESDF conversion introduces quasi-Euclidean approximations. FIESTA~\cite{hanFIESTAFastIncremental2019} eliminated this stage by updating ESDFs directly from occupancy grids, providing a true Euclidean distance map. More recently, neural and continuous representations such as LGSDF~\cite{yueLGSDFContinualGlobal2025} have traded computational overhead for smoother, globally consistent fields.
 
Projective TSDF integration, as used by KinectFusion and nvblox~\cite{millaneNvbloxGPUAcceleratedIncremental2024}, builds a range image from each depth frame and updates voxels by projecting them into the nearest angular bin. This formulation is correct for dense RGB-D and fixed-beam LiDAR sensors, where every bin corresponds to an emitted ray; however, Voxfield~\cite{panVoxfieldNonProjectiveSigned2022} demonstrated that even under these conditions, projective integration systematically overestimates signed distances at large incidence angles, proposing non-projective raycasting as a remedy. For non-repetitive sparse sensors such as the Livox Mid-360, the failure mode is more fundamental: single-frame point clouds are sparse and randomly distributed~\cite{wangSFPNetSparseFocal2025}, so any raycasting approach propagates free-space estimates through the majority of the volume each frame, progressively eroding thin structures --- a failure mode Voxfield's non-projective raycasting does not address.
 
Existing reachability and visibility methods, surveyed next, similarly fall short of a live, scene-level accessibility field.

\subsection{Reachability and Visibility}
\label{subsec:reach_planning}
The most directly related perception output to our own is point-level visibility. Katz \textit{et al.}~\shortcite{katzDirectVisibilityPoint2007} proposed Hidden Point Removal (HPR), a method for determining visible points from a given viewpoint without surface reconstruction, while more recent work~\cite{wangNeuralVisibilityPoint2025} achieved significant performance improvements through a learned approach. Visibility is, however, a strictly weaker property than accessibility: a point visible from a given direction may still be inaccessible because the tool shaft geometry collides with surrounding structure along the approach path, or may lie on a surface face never observed from an approachable side. Our method addresses both failure modes through explicit tool-kernel collision checking and observation-side filtering of candidate approach directions.

Visual affordance learning addresses a related question: 3D AffordanceNet~\cite{deng3DAffordanceNetBenchmark2021} provides per-point affordance annotations across 23k object shapes, labelling which interactions each surface region supports. These labels are semantic and learned per object category, describing what an object affords rather than whether a specific tool can physically reach a given surface, and they transfer to neither novel tool geometries nor unstructured outdoor scenes. Point-specific feasibility scoring has also been explored in robotic grasping. PointNetGPD~\cite{liangPointNetGPDDetectingGrasp2019} and 6-DOF GraspNet~\cite{mousavian6DOFGraspNetVariational2019} predict grasp quality for local point clouds and gripper configurations, and GraspNet-1Billion~\cite{fangGraspNet1BillionLargeScaleBenchmark2020} provides a large-scale benchmark for such evaluations.
 These methods are the closest in spirit to our own, but are object-centric, operating on complete segmented object scans rather than unstructured scene-level point clouds. Tool geometry is encoded through training rather than explicit geometric checks, limiting generalisation to novel tool shapes without retraining. Their benchmarks do not transfer for the same reason: GraspNet-1Billion's indoor, gripper-closure grasp labels do not define accessibility for a tool shaft approaching through outdoor canopy, so we validate instead against an independently implemented mesh-collision ground truth on our own target geometry (\Cref{subsec:synthetic}).

Our target application of aerial interaction introduces constraints that none of the above address. Although robotic pruning has been studied extensively for ground-based platforms in orchards and vineyards~\cite{navoneAutonomousRoboticPruning2025}, their perception pipelines focus on branch detection and localisation in structured crops. Aerial platforms present an emerging approach for environmental interaction~\cite{linDeepLearningBasedDepth2024}, but introduce additional perceptual challenges: unbounded outdoor environments, live LiDAR from a moving sensor, and tool approach paths constrained by rotor clearance.

\section{Proposed Method}
\label{sec:method}

Given a LiDAR stream and a tool mesh $\mathcal{T}$, we compute a per-point accessibility label $A(\mathbf{s}) \in \{\textrm{accessible}, \textrm{blocked}, \textrm{unknown}\}$ for every point in the accumulated, incrementally growing point cloud, updated as new frames arrive. Sensor poses are assumed to come from an upstream LiDAR-inertial odometry system~\cite{zhangLOAMLidarOdometry2014,shanLIOSAMTightlycoupledLidar2020}; residual pose error is absorbed by the confidence-weighted TSDF rather than corrected explicitly.

\begin{figure}[h]
  \centering
  \includegraphics[width=0.8\linewidth]{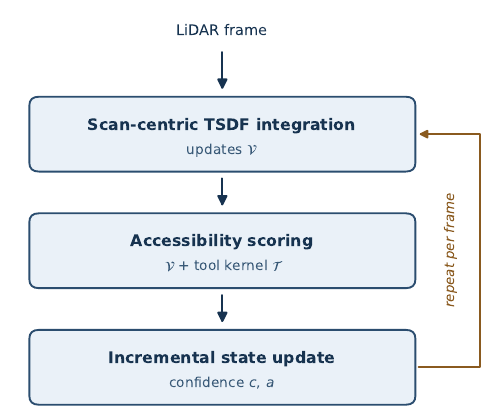}
  \caption{\textbf{System pipeline.} Each incoming LiDAR frame is integrated into the TSDF $\mathcal{V}$, scored against candidate approach directions using the tool kernel $\mathcal{T}$, and the resulting accessibility state is updated incrementally before the next frame arrives.}
  \label{fig:pipeline}
\end{figure}

Each incoming frame is accumulated into a voxel-deduplicated point cloud, from which the TSDF voxel grid $\mathcal{V}$ is built incrementally (\Cref{subsec:scan_tsdf}). Accessibility labels are attached to the point cloud; the distances used to compute these come from $\mathcal{V}$. The three stages run once per frame (\Cref{fig:pipeline}).

\subsection{Scan-centric TSDF}
\label{subsec:scan_tsdf}

We represent the scene as a TSDF on a regular voxel grid with voxel size $\ell$ and truncation distance $\tau$. $\tau$ is selected based on each tool's physical size, subject to a floor of $4\ell$ so the truncation band remains representable at the chosen resolution. Arm-path samples (\Cref{subsec:access_score}) are then placed at fixed multiples of $\tau$ along the tool's approach ray, so they fall at distances that actually clear the tool body. Following the same TSDF running-average formulation as KinectFusion~\cite{newcombeKinectFusionRealtimeDense2011}, each voxel stores a running sum $S$ and observation count $w$. For each new scan point $\mathbf{p}_i$ observed from sensor origin $\mathbf{v}$, every voxel centre $\mathbf{x}$ within the update neighbourhood has its running sum and count updated as follows:

\begin{equation}
d = \|\mathbf{p}_i - \mathbf{v}\| - \|\mathbf{x} - \mathbf{v}\|,
    \quad \text{clamped to } [0, \tau]
\label{eq:tsdf_dist}
\end{equation}

Clamping to $[0,\tau]$ leaves the stored field non-negative everywhere, so it is not signed in the usual sense: voxels on or behind a surface both read zero, and only the free-space side carries a distance. We keep the TSDF terminology because the update rule, truncation and running average follow the standard formulation, and the accessibility scorer requires only clearance magnitude, never the inside/outside sign.

\begin{equation}
S \leftarrow S + d, \qquad w \leftarrow w + 1
\label{eq:tsdf_accum}
\end{equation}

$S$ and $w$ accumulate across every observation of that voxel, so more observations produce a more reliable running mean.

\begin{equation}
  \mathrm{TSDF}(\mathbf{x}) =
    \begin{cases} S / w & w > 0 \\ \tau & w = 0 \text{ (free)} \end{cases}
  \label{eq:tsdf_query}
\end{equation}

We define $w = 0$ as the unknown state; unobserved voxels read as free for tool kernel checks but occupied for the arm path checks, a convention justified in \Cref{subsec:access_score}.

\textbf{Projective integration.}
As discussed in \Cref{sec:background}, projective TSDF integration breaks down for non-repetitive sparse sensors: in our outdoor dataset, a single 100\,\unit{\milli\second} scan using our Livox Mid-360~\cite{livoxMid360manual} leaves approximately 96\% of \ang{0.25}$\times$\ang{0.25} angular bins without returns, and even at the coarser $1024 \times 64$ resolution of a fixed-beam sensor (Ouster OS1~\cite{ousterOS1datasheet}), 81\% remain empty. At this sparsity, projective lookup assigns free-space distances to voxels on thin structures from rays they do not lie on, progressively eroding the distance field on which accessibility scoring depends.

\begin{figure}[H]
\centering
\includegraphics[width=\linewidth]{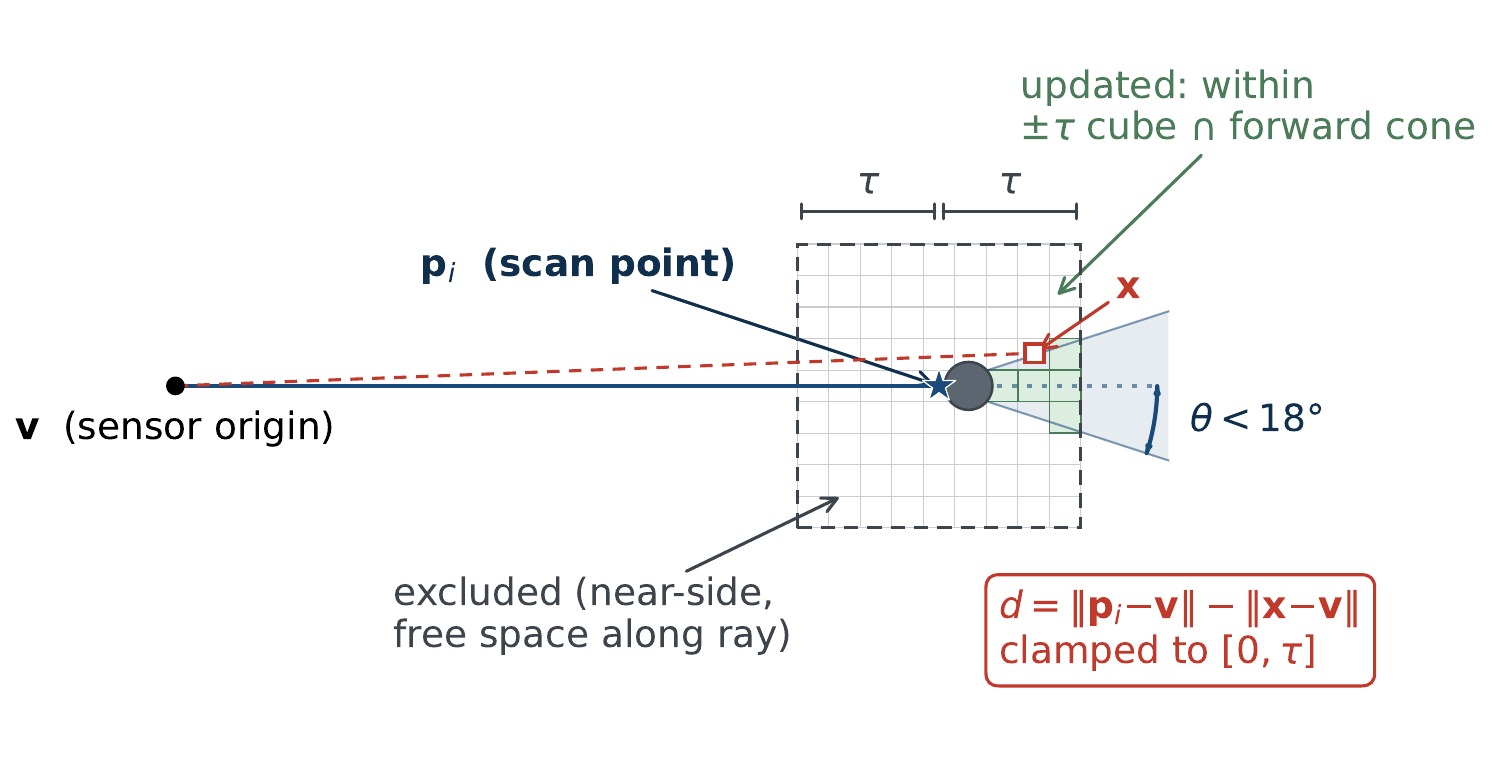}
\caption{\textbf{Scan-centric TSDF update.} Voxels within the $\pm\tau$ cube and forward cone around scan point $\mathbf{p}_i$ are updated (green); one such voxel centre $\mathbf{x}$ is highlighted, at distance $d$ from $\mathbf{p}_i$ per \Cref{eq:tsdf_dist}. Near-side voxels are excluded as free space.}
\label{fig:scan_centric}
\end{figure}

\textbf{Scan-centric integration.}
To avoid this failure mode, we update only the voxels in the immediate neighbourhood of each observed return: every voxel within the $\pm\tau$ cube and forward cone around each scan point is updated, while voxels on the near side are excluded as free space along the ray to the sensor, as shown in \Cref{fig:scan_centric}. 

Note that clamping does not guarantee the mean remains above the collision threshold $r_c$ (\Cref{subsec:access_score}) when many zero contributions accumulate from near-grazing observations; in practice, the confidence threshold $c < 0.2$ (\Cref{subsec:updateconf}) filters such points before scoring. Voxels outside the $\pm\tau$ neighbourhood of any return are never written, retaining their default value of $\tau$ (free).

\subsection{Sliding-Window TSDF}
\label{subsec:sliding}

For unbounded outdoor scenes, a fixed grid would either exhaust memory or require a coarse voxel resolution. To address this, we use a sliding window of size $(L_x, L_y, L_z)$ centred on the sensor viewpoint. When the sensor moves more than 30\% of the window dimension along any axis, the grid shifts: we copy TSDF data in the overlapping region to the correct offset, discard voxels leaving the window, and initialise new cells to $w = 0$ (unknown). This approach bounds memory use to $\mathcal{O}(L_x L_y L_z / \ell^3)$. Points leaving the window keep their last label and are re-scored once their re-initialised voxels regain confidence (\Cref{subsec:updateconf}).

\subsection{Accessibility scoring}
\label{subsec:access_score}

Given the distance field, we evaluate per-point surface accessibility by placing a precomputed tool-geometry kernel at each surface point and testing candidate approach directions for collisions.

\textbf{Approach directions.}
We sample $D = 180$ candidate approach directions uniformly over $[0, 2\pi)$ in yaw, with a fixed pitch consistent with our target platform's horizontal-approach constraint, then reduce this set with two filters. First, a normal filter, analogous to C-space accessibility cones~\cite{spyridiAccessibilityAnalysisAutomatic1990,limaiemGeneralMethodAccessibility1997}, retains only directions $\mathbf{u}$ within a \ang{70} half-cone of the outward surface normal $\mathbf{n}$, a threshold chosen to allow oblique side-approaches while excluding directions that would require the tool to approach through the surface itself. The normal is estimated via PCA on the $k = 10$ nearest neighbours and oriented to face the mean of all historical sensor positions recorded for that surface point $\mathbf{s}$.

\begin{equation}
-(\mathbf{n} \cdot \mathbf{u}) \geq \cos(\ang{70})
\label{eq:normal}
\end{equation}

Second, an observation-side filter requires positive alignment with at least one of up to eight stored historical viewpoints $\mathbf{v}_m$; directions approached only from the back of a surface are excluded, preventing false accessibility labels on unobserved surface faces.

\begin{equation}
\exists\, \mathbf{v}_m : \frac{\mathbf{v}_m - \mathbf{s}}{\|\mathbf{v}_m - \mathbf{s}\|} \cdot \mathbf{u} > 0.25
\label{eq:obsside}
\end{equation}

\textbf{Tool kernel.}
We represent tool geometries as a point cloud $\mathcal{T} = \{\mathbf{t}_k\}_{k=1}^{M}$ sampled from the tool's 3D model, with the approach axis aligned to $+x$ and the interaction point (e.g., a probe tip) at the origin. We precompute a set of yaw-rotated kernels: for each yaw angle $\psi_j$, $j = 1,\ldots,D$, sampled uniformly over $[0, 2\pi)$, the rotated kernel is $\mathcal{T}_j = \{\mathbf{t}_{k, j}\}_{k=1}^{M} $ where $\{\mathbf{t}_{k,j}\} = R_z(\psi_j)\,\mathbf{t}_k$ and $R_z(\psi_j)$ denotes rotation about the vertical $(z)$ axis. Our target aerial platform constrains tool approach to horizontal, making yaw the sole degree of freedom; however, implementing full spherical sampling is straightforward for less constrained tools. All $D$ kernels are stored as a single tensor of shape $\mathbb{R}^{M \times 3 \times D}$ in GPU memory, allowing direct indexing without runtime rotation.

To score a candidate direction $\mathbf{u}$ at surface point $\mathbf{s}$, the kernel points $\mathbf{s} + \mathbf{t}_{k,j}$ are queried against the TSDF. A direction is accessible if the collision count (kernel points with TSDF $< r_c$) falls below the threshold $n_c$. Both $r_c$ and the self-exclusion radius $r_s$ scale with the tool and voxel resolution in use; values for each experiment are given in \Cref{subsec:synthetic}.

\textbf{Arm path checking.}
The tool kernel captures end-effector geometry but not the approach corridor behind it. We sample the TSDF at three fixed distances $\{0.6\tau, 1.25\tau, 2.25\tau\}$ along the approach ray from the surface; if any sample falls below $r_c$, the direction is blocked immediately. To support this check, we maintain a boolean free-space tensor updated during TSDF integration: for each return, voxels in the band $(\tau, 2.25\tau]$ along the approach ray are marked ray-cleared. This band is append-only and does not modify TSDF values. Unobserved voxels ($w = 0$) that are not ray-cleared are treated as occupied: the approach corridor could extend into unmapped space, and treating unknown regions as free could permit approaches through unobserved obstacles.

\textbf{Self-exclusion.}
Tool points within radius $r_s$ of the surface point are excluded from the collision check, preventing the surface voxel from blocking all directions and allowing the tool tip to make direct contact.

\subsection{Incremental Updates and Confidence}
\label{subsec:updateconf}

Full re-evaluation at each frame becomes expensive as the map grows; for deployment on a robot with limited edge compute, this is unacceptable. To address this, we maintain a per-point accessibility state and trigger re-evaluation only under three conditions: (i) the TSDF at a point's voxel changes by more than $2\ell$ between consecutive frames; (ii) any voxel within neighbourhood radius $r_u = 3\ell$ changes, determined by sampling over at most 256 changed voxels per frame to bound query cost; (iii) the sensor moves more than $0.1\,\unit{\metre}$, which re-queues all points currently within the sliding window that are classified as inaccessible, since the observation-side filter depends on accumulated viewpoints. New points are always evaluated on first observation. An adaptive per-frame budget cap limits evaluations and defers overflow to a backlog consumed between LiDAR frames.

Each surface point carries a TSDF confidence $c = w / (w + k_c)$, $k_c = 5$, a count-based function that rises from 0 toward 1 as the number of TSDF observations $w$ accumulates, normalised so it can combine multiplicatively with the accessibility confidence $a$ below rather than operating on unbounded observation counts directly; at $w = k_c$ observations, $c = 0.5$.  Points with $c < 0.2$ are skipped during scoring, preserving their last known score rather than resetting to unknown. Per-direction scoring is the raw collision count (kernel points with TSDF $< r_c$), and accessible classification requires a score below $n_c$. A point is classified as \textit{confident-accessible} when its best stored direction score is below $n_c$ and its combined confidence $c \cdot a > 0.7$. A direction found clear is never unilaterally blocked by a single conflicting re-evaluation, guarding against a single noisy or occluded frame; instead, conflicting evidence accumulates through an accessibility confidence $a$ (initialised to 1), which decays by a factor of 0.75 per conflicting evaluation and recovers by $0.15 \cdot (1 - a)$ per confirming evaluation: for a well-observed point, two consecutive blocking evaluations remove it from the accessible set, while a single outlier recovers. Stored best-point scores are non-increasing: re-evaluation updates the stored score only when a new score is strictly lower.

\subsection{GPU implementation}
\label{subsec:gpu}

The pipeline runs entirely on GPU in three stages: scan-centric integration (including sliding-window management), accessibility
scoring, and incremental state updates. All stages must complete within a single LiDAR frame period ($\sim$100\,\unit{\milli\second}).

\textbf{TSDF integration.}
The integration kernel launches one thread per (scan point, voxel offset) pair, processing the $(2\lceil\tau/\ell\rceil + 1)^3$ neighbourhood cube of each return in parallel. Scan points are processed in fixed-size chunks to bound VRAM usage. Voxel observation counts are stored as \texttt{float16}, capped at 2000 observations with proportional rescaling of the running sum to prevent saturation.

\textbf{Accessibility scoring.}
The tool kernel tensor and all $D$ yaw-rotated kernels are precomputed once at initialisation and held in GPU memory, without runtime rotation. A coarse-to-fine direction search --- a $\ang{20}$ coarse pass followed by a $\pm\ang{25}$ fine window around the best direction of the nearest resolved neighbour within $0.3\,\unit{\metre}$ --- reduces the mean number of full kernel evaluations per frame.

\textbf{Performance.}
On an NVIDIA RTX 4090, the complete pipeline processes a $1\,000\,000$+ point map at $1\,\unit{\centi\metre}$ voxel size in under $30\,\unit{\milli\second}$ per LiDAR frame on average.

\pagebreak

\section{Experiments and results}
\label{sec:results}

As per-point accessibility analysis is a novel field, there are no pre-existing ground-truth datasets. Physical testing of surfaces to generate this ground truth would be prohibitively slow, so we design three experiments: (i) an initial evaluation on synthetically generated simple objects, with exhaustively validated ground truth; (ii) a larger-scale evaluation on synthetic pine tree models; and (iii) real-world testing on LiDAR data from outdoor tree scans. The mapping-backend and clamping ablations (\Cref{subsec:synthetic}) isolate the contribution of scan-centric, positive-clamped integration, and the visibility comparison (\Cref{subsec:vis_comp}) isolates the accessibility field against visibility and point-cloud baselines.

\subsection{Synthetic Evaluation}
\label{subsec:synthetic}

\begin{figure*}[t]
    \centering
    \begin{subfigure}[b]{0.19\textwidth}
        \centering
        {\setlength{\fboxsep}{4pt}%
        \colorbox{blue!10}{\includegraphics[width=\dimexpr\linewidth-8pt\relax]{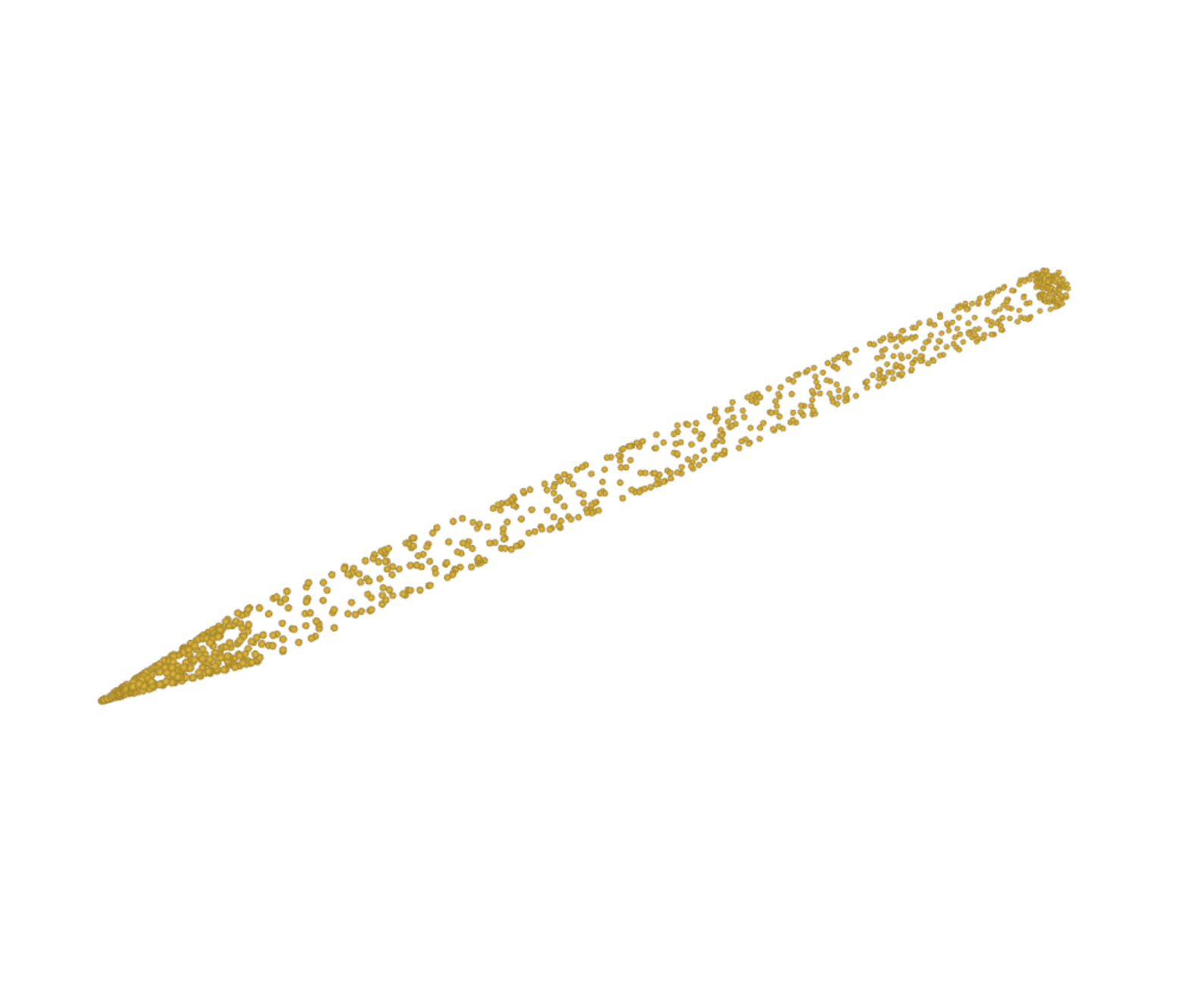}}}
        \caption{}
        \label{fig:tool-mesh}
    \end{subfigure}
    \hfill
    \begin{subfigure}[b]{0.19\textwidth}
        \includegraphics[width=\linewidth]{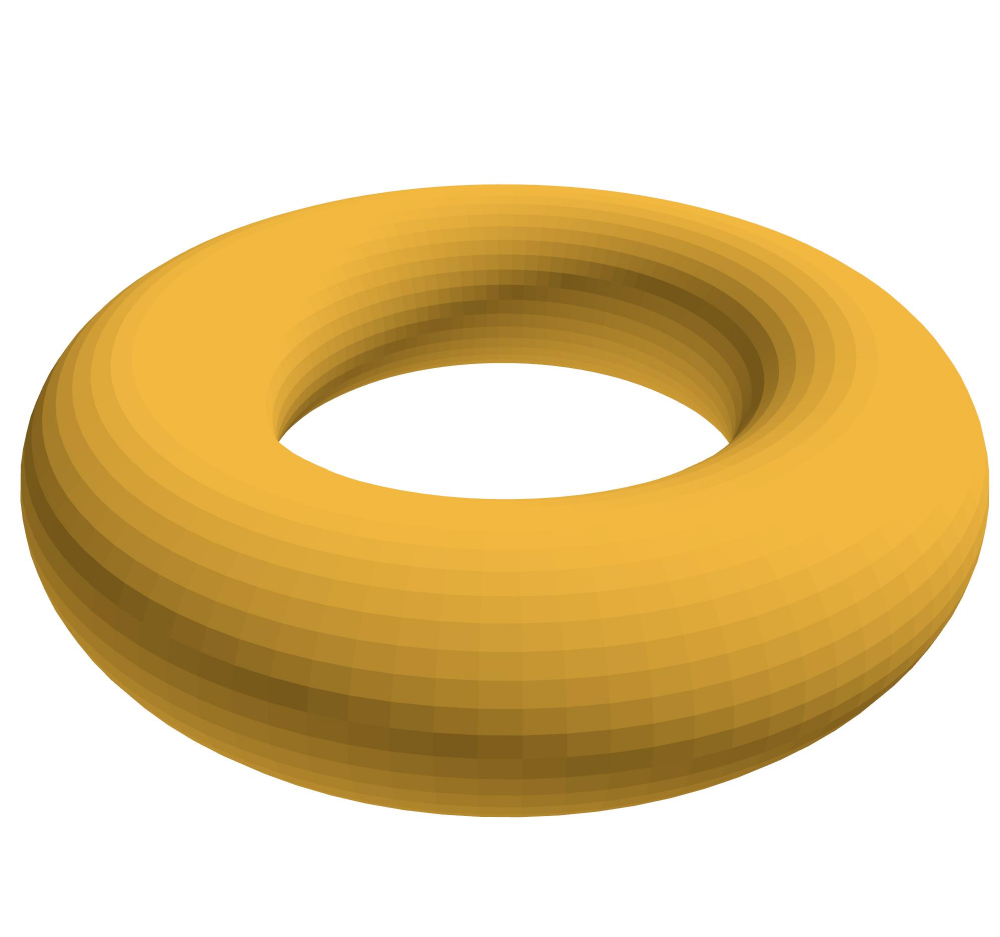}
        \caption{}
        \label{fig:torus-mesh}
    \end{subfigure}
    \hfill
    \begin{subfigure}[b]{0.19\textwidth}
        \includegraphics[width=\linewidth]{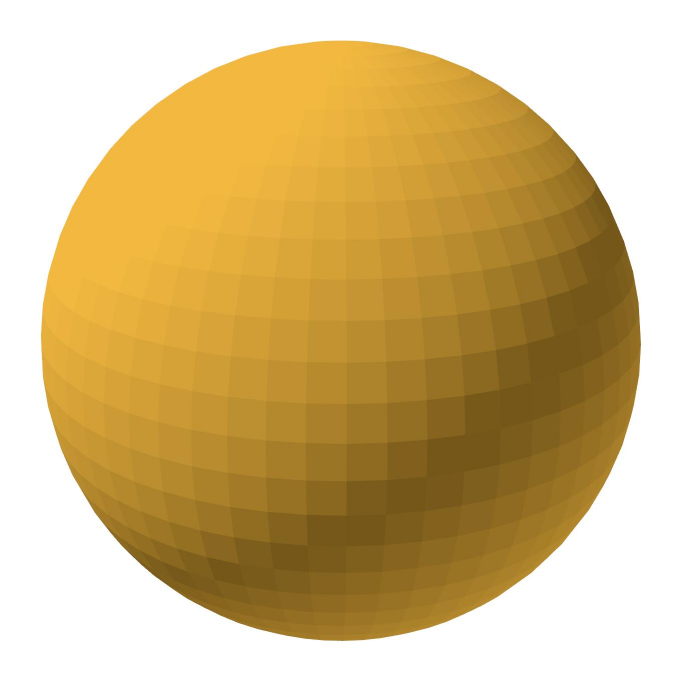}
        \caption{}
        \label{fig:sphere-mesh}
    \end{subfigure}
    \hfill
    \begin{subfigure}[b]{0.19\textwidth}
        \includegraphics[width=\linewidth]{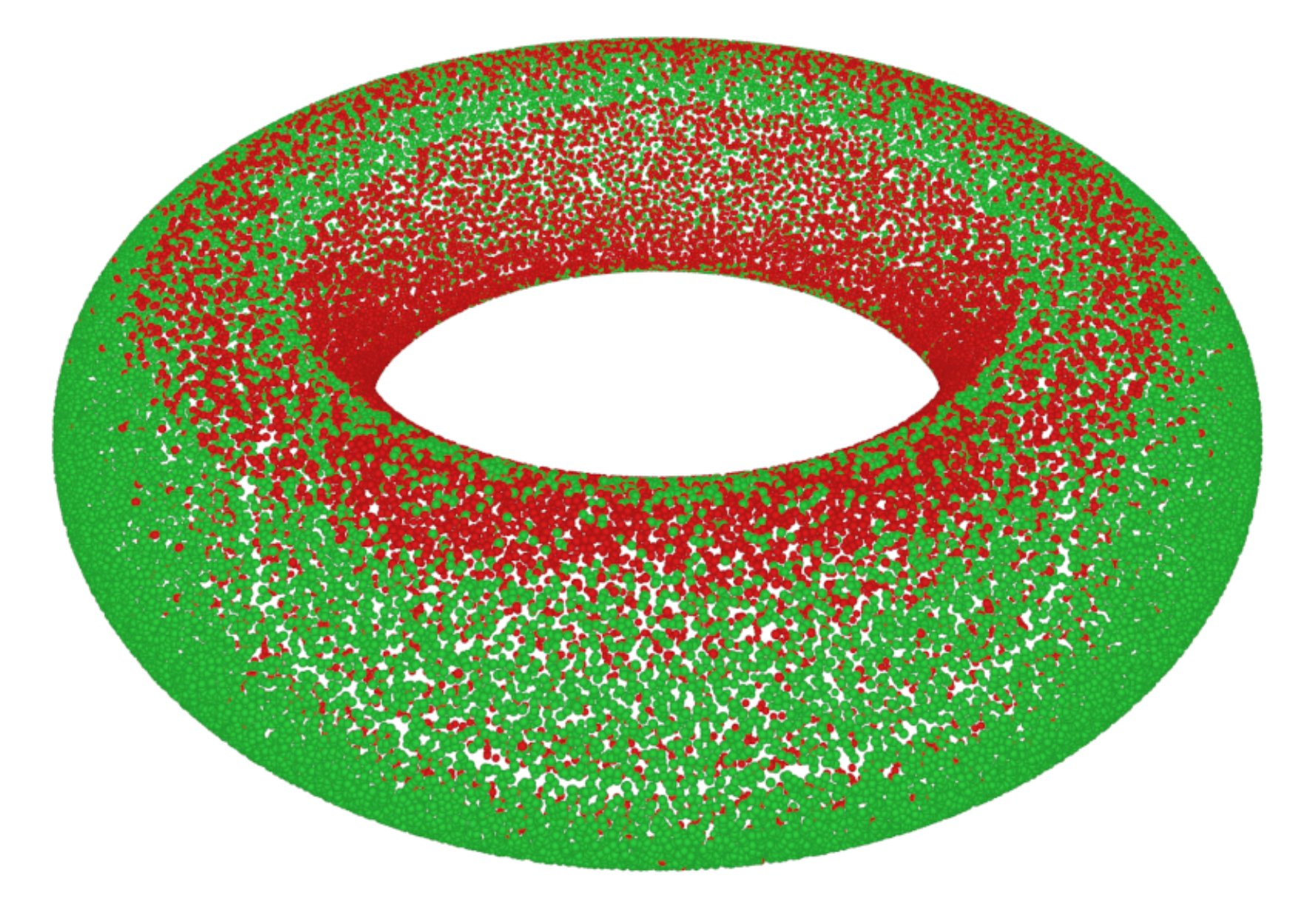}
        \caption{}
        \label{fig:torus-access}
    \end{subfigure}
    \hfill
    \begin{subfigure}[b]{0.19\textwidth}
        \includegraphics[width=\linewidth]{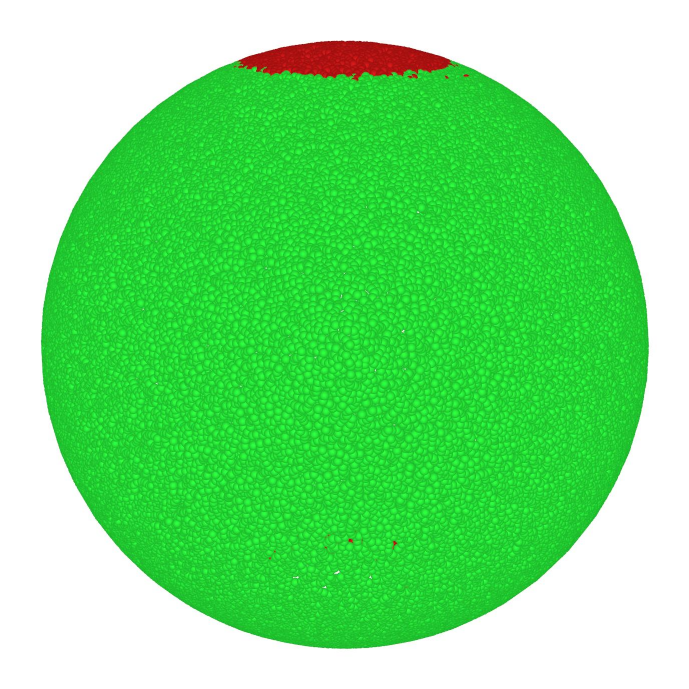}
        \caption{}
        \label{fig:sphere-access}
    \end{subfigure}

    \caption{\textbf{Tool geometry and synthetic evaluation scenes.} Pencil tool (a), torus (b) and sphere (c) meshes, with the corresponding per-point accessibility results at 1\,\unit{mm} voxel size (d, e), tool rotation limited to yaw only. The box and wall scenes are omitted: both are entirely accessible, so their result renders are uniformly green.}
    \label{fig:synthetic-scenes}
\end{figure*}

\textbf{Ground truth.}
Accessibility ground truth is exhaustively computed for all $D{=}180$ approach directions against an exact SDF of each synthetic object, using a pencil-shaped tool geometry (\Cref{fig:tool-mesh}). As outlined in \Cref{subsec:access_score}, for each surface point a direction is accepted if (i) it lies within a \ang{70} half-cone of the outward normal, (ii) three fixed arm-path samples at $\{0.6\tau,\,1.25\tau,\,2.25\tau\}$ do not intersect the mesh, and (iii) the tool-kernel collision count falls below the threshold $n_c{=}5$. The pencil tool uses $\tau = 15\,\unit{\milli\metre}$, $r_c = 1\,\unit{\milli\metre}$ and $r_s = 5\,\unit{\milli\metre}$ at this resolution. Scans are simulated with Open3D raycasting using a Halton ray pattern that approximates the Livox Mid-360 at 10\,\unit{\hertz}, and metrics are reported for points with TSDF confidence $c > 0.2$. \Cref{tab:oracle} shows results at 1\,\unit{\milli\metre} voxel size.

\begin{table}[h]
\centering
\caption{\textbf{Analytical ground truth accessibility evaluation at 1\,mm voxel size} (scan-centric, positive-clamped TSDF).}
\label{tab:oracle}
\begin{tabular}{lrrrrr}
\toprule
Scene & $N$ & GT access. & Prec & Rec & F1 \\
\midrule
Torus  & 42,790  & 53.6\,\%  & 83.2  & 100.0 & 90.8 \\
Sphere & 62,023  & 96.5\,\%  & 100.0 & 100.0 & 100.0 \\
Box    & 179,812 & 100.0\,\% & 100.0 & 93.3  & 96.6 \\
Wall   & 44,753  & 100.0\,\% & 100.0 & 100.0 & 100.0 \\
\bottomrule
\end{tabular}
\end{table}

Three of four scenes achieve F1\,$\geq$\,96.6, with zero false negatives across the sphere and wall, and only four false negatives across the full torus surface. False positives on the torus (16.8\,\% of accessible predictions) are confined to a narrow strip at the transition between the accessible outer tube and the inaccessible inner tube, where the 1\,\unit{\milli\metre} TSDF cannot perfectly represent the curved accessibility boundary. The box shows a 6.7\,\% false-negative rate at edges and corners where the arm-path check clips the geometry. Unlike the torus false positives, this rate is stable across voxel sizes (\Cref{tab:voxel}): $\tau$, and hence the arm-path sample distances, are held fixed, so the same extremities are clipped at every resolution. The system's main failure mode is thus false-negative accessibility at geometric extremities, an appropriate behaviour given that a voxel's accessibility may be critical for interaction planning. To confirm this ground truth is not itself an artefact of the analytical SDF definition, we additionally validate it against an independently implemented mesh-collision ground truth on the two curved-boundary scenes: agreement reaches F1\,=\,99.96 on the torus and F1\,=\,100.0 on the sphere, confirming the two independently computed ground truths are not simply restating the same rule.

\begin{table}[h]
\centering
\caption{\textbf{Voxel-size and rotation sensitivity.} Torus is the only scene with non-trivial voxel-size sensitivity: box F1 varies only between 96.6 and 96.8 across 1--5\,mm, and the wall is exact at every resolution, so those rows are omitted.}
\label{tab:voxel}
\begin{tabular}{llrrrr}
\toprule
Scene & Param & $N$ & Prec & Rec & F1 \\
\midrule
Torus & 1\,mm & 42,790  & 83.2  & 100.0 & 90.8 \\
      & 2\,mm & 14,266  & 74.1  & 100.0 & 85.1 \\
      & 5\,mm & 2,437   & 76.5  & 100.0 & 86.7 \\
\midrule
Box   & \ang{0}  & 179,812 & 100.0 & 93.3 & 96.6 \\
(1\,mm)      & \ang{22} & 173,073 & 100.0 & 82.2 & 90.2 \\
\bottomrule
\end{tabular}
\end{table}

\textbf{Voxel size.}
\Cref{tab:voxel} shows the effect of voxel size and rotation on accessibility labelling. On the torus, recall remains 100\,\% at all resolutions; precision degrades from 83.2\,\% at 1\,\unit{\milli\metre} to 74.1\,\% at 2\,\unit{\milli\metre} as the TSDF representation of the curved tube becomes coarser, before a partial recovery at 5\,\unit{\milli\metre} (F1\,=\,86.7) that is likely sampling noise given the small point count at that resolution (2,437 points). Flat-surface scenes are insensitive to voxel size because their accessibility boundaries are axis-aligned at \ang{0} rotation; rotating the box \ang{22} off-axis, however, reduces recall from 93.3\,\% to 82.2\,\% (F1 96.6\,$\rightarrow$\,90.2) at $1\,\unit{\milli\metre}$, the worst of the three off-axis rotations tested, confirming that axis-alignment, not voxel size, was responsible for the box's apparent insensitivity.

\textbf{Mapping backend.}
The closest published alternative to our integration scheme is projective TSDF fusion as implemented by nvblox~\cite{millaneNvbloxGPUAcceleratedIncremental2024}. To compare on equal terms, we substitute a projective integration following that formulation into our pipeline and hold the accessibility stage fixed; the only variable is how we build the distance field. It collapses the sphere's confident surface points by 91\,\% (62,023 $\rightarrow$ 5,676), rendering all remaining points inaccessible (F1\,=\,0), and collapses torus recall to 20.4\,\% (F1\,=\,33.9) as erosion preferentially targets the accessible outer surface. Positive clamping is the second design choice from \Cref{subsec:scan_tsdf}. Evaluated at $2\,\unit{\milli\metre}$, removing it trades a different failure: sphere recall falls to 56.3\,\% (F1\,=\,72.1), with 10,009 accessible points incorrectly marked blocked, while the torus shows a mixed effect (precision improves from 74.1\,\% to 100\,\%, at a cost of 1,413 new false negatives). Positive clamping is the correct choice for this application: it accepts boundary imprecision on curved geometry in exchange for zero false negatives on unambiguous surfaces, the safer failure mode given a downstream planner can reject candidates via physical collision checking.

\subsection{Visibility Comparison}
\label{subsec:vis_comp}

We quantify the gap between visibility and accessibility using HPR~\cite{katzDirectVisibilityPoint2007}: HPR is applied from every simulated sensor viewpoint per scene, and a point counts as visible under this baseline if it is visible from at least one viewpoint, the union across all viewpoints, not a single view. HPR has no notion of tool geometry at all, so we add a second, ablative baseline to separate the contribution of visibility from that of the tool. It runs our own candidate-direction sweep, cone filter and arm-path check unchanged, but answers each collision query with a nearest-neighbour distance to the accumulated point cloud instead of a TSDF lookup. It therefore sees exactly the data our method sees, with no access to the ground-truth mesh, and isolates what the volumetric representation contributes.

\textbf{Primitive scenes.}
\Cref{tab:hpr} reports precision, recall and F1 against our synthetic ground truth for all three methods, applying our pencil-shaped tool throughout. HPR achieves 100\,\% recall on every scene, a trivial consequence of classifying all visible points as accessible once sufficient viewpoints have accumulated, making precision the discriminating metric. Only the torus carries a substantial inaccessible region: 46.4\,\% of its surface lies on the inner ring, against 3.5\,\% rear-facing points on the sphere and none at all on the box and wall. It is therefore the only scene of the four on which these metrics separate the three methods.

On the torus, HPR reaches a precision of 53.6\,\% (F1\,=\,69.8), having no mechanism to detect that the tool shaft would collide with the surrounding torus geometry along the approach path. The point-cloud baseline, which does model tool geometry, recovers much of that gap at 68.8\,\% precision (F1\,=\,81.5), and our method reaches 83.2\,\% (F1\,=\,90.8). The ordering separates the two contributions: explicit tool-kernel collision checking accounts for most of the improvement over visibility alone, and the TSDF representation, which a nearest-neighbour query over the raw point cloud does not reproduce, for the rest. On the remaining scenes precision is bounded at 100\,\% for any method that labels generously, and all three are correspondingly near-perfect; our method's lower box recall (93.3\,\%) is the arm-path edge clipping of \Cref{subsec:synthetic}.

\begin{table}[t]
\centering
\caption{\textbf{Visibility and collision baselines vs.\ ours} at $1\,\unit{\milli\metre}$ voxels (confidence $>0.2$). \emph{HPR}: visible from at least one viewpoint. \emph{Point cloud}: our direction sweep and arm-path check, with collision queries answered by nearest-neighbour distance to the accumulated point cloud rather than the TSDF, and no mesh access.}
\label{tab:hpr}
\begin{tabular}{llrrr}
\toprule
Scene & Method & Prec & Rec & F1 \\
\midrule
Torus  & HPR         & 53.6 & 100.0 & 69.8 \\
       & Point cloud & 68.8 & 100.0 & 81.5 \\
       & Ours        & \textbf{83.2} & 100.0 & \textbf{90.8} \\
\midrule
Sphere & HPR         & 96.5 & 100.0 & 98.2 \\
       & Point cloud & 100.0 & 100.0 & 100.0 \\
       & Ours        & 100.0 & 100.0 & 100.0 \\
\midrule
Box    & HPR         & 100.0 & 100.0 & 100.0 \\
       & Point cloud & 100.0 & 100.0 & 100.0 \\
       & Ours        & 100.0 & 93.3 & 96.6 \\
\midrule
Wall   & HPR         & 100.0 & 100.0 & 100.0 \\
       & Point cloud & 100.0 & 100.0 & 100.0 \\
       & Ours        & 100.0 & 100.0 & 100.0 \\
\bottomrule
\end{tabular}
\end{table}

\begin{figure}[t]
    \centering
    \begin{subfigure}[b]{0.49\linewidth}
        \centering
        \includegraphics[width=\linewidth, trim=8pt 8pt 8pt 8pt, clip]{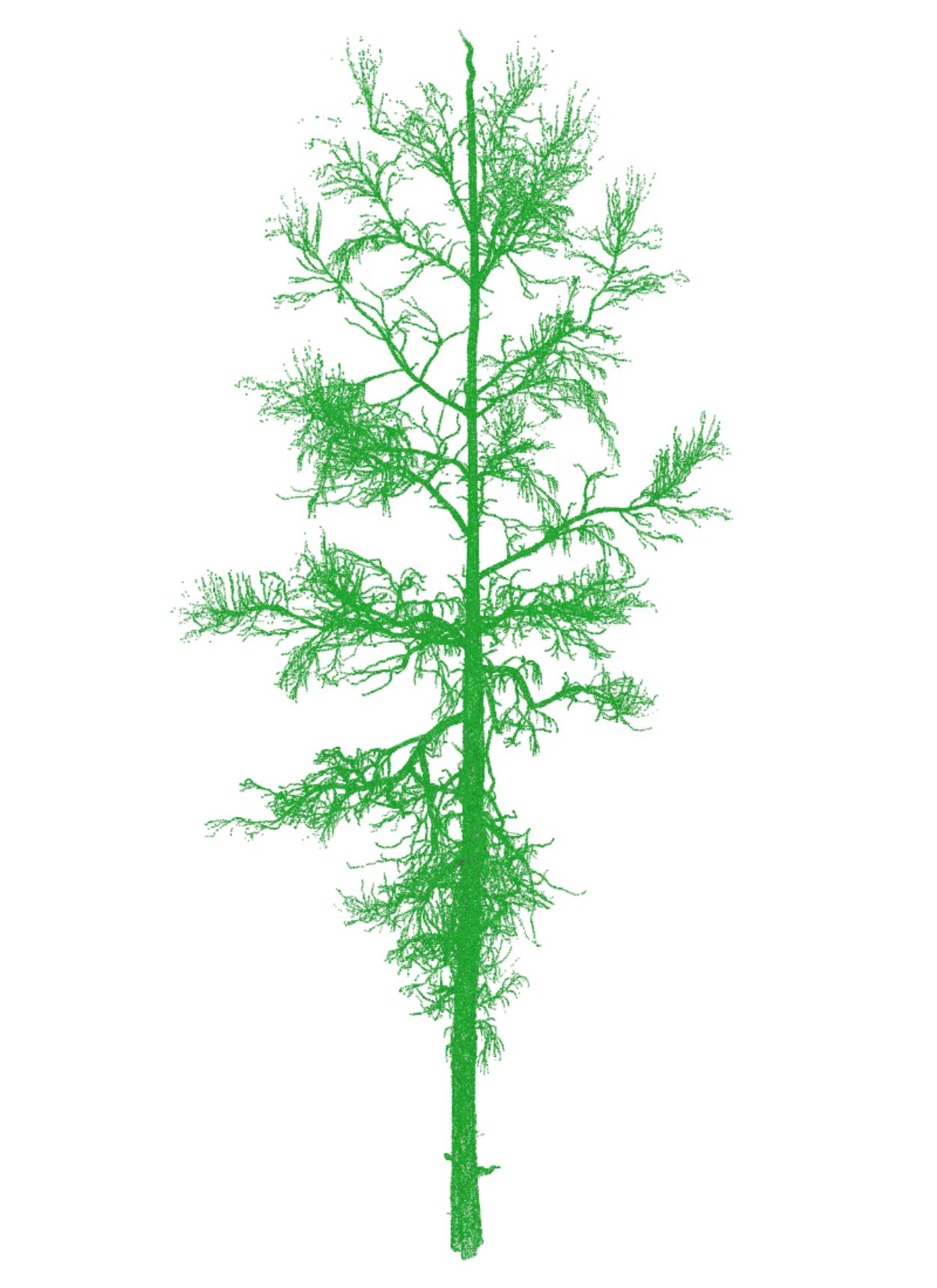}
        \caption{}
        \label{fig:HPR_pinus}
    \end{subfigure}
    \hfill
    \begin{subfigure}[b]{0.49\linewidth}
        \centering
        \includegraphics[width=\linewidth, trim=8pt 8pt 8pt 8pt, clip]{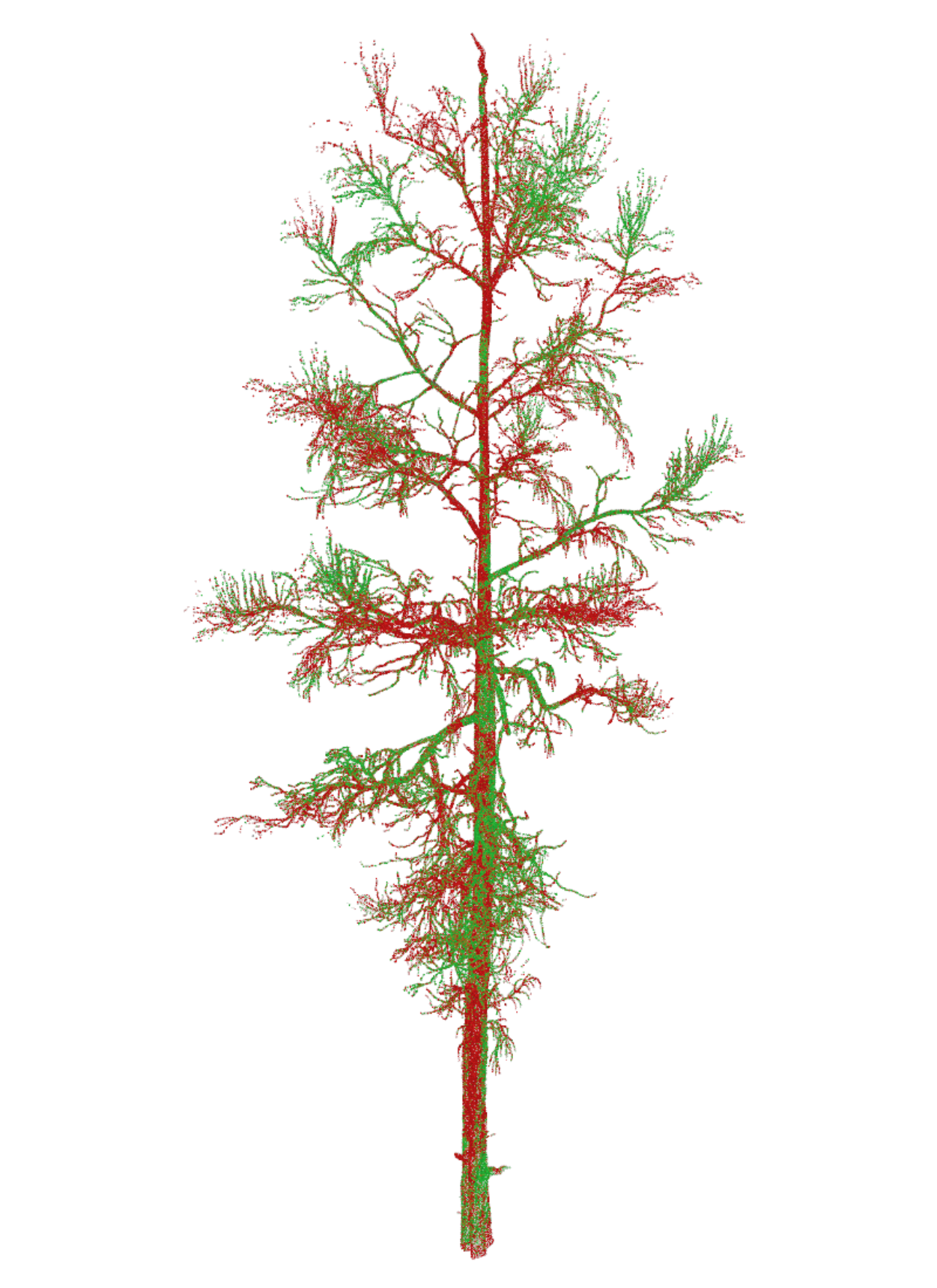}
        \caption{}
        \label{fig:acc_pinus}
    \end{subfigure}
    \caption{\textbf{Visibility vs.\ accessibility for synthetic Pinus radiata.} (a) HPR labels 95.0\,\% of surface points as visible (green). (b) Our method labels 38.1\,\% as accessible; surface points blocked by surrounding geometry are marked inaccessible (red).}
    \label{fig:synth_pinus}
\end{figure}

\textbf{Synthetic \textit{Pinus radiata}.}
We also extend this comparison to a mature-scale synthetic \textit{Pinus radiata} tree, using a tool-geometry kernel representing our UAV pruning platform. HPR is run for every fifth trajectory viewpoint over 487,655 surface points accumulated during the simulated LiDAR scan, at a voxel size of $10\,\unit{\milli\metre}$ with $\tau = 0.30\,\unit{\metre}$, $r_c = 2\,\unit{\centi\metre}$ and $r_s = 5\,\unit{\centi\metre}$. HPR labels 95.0\,\% of surface points as visible; our method labels 38.1\,\%. This 56.8 percentage-point gap represents points visible from the sensor trajectory that our method labels inaccessible because surrounding branch geometry obstructs an approach. As a result, HPR produces a near-uniformly green canopy as shown in \Cref{fig:HPR_pinus}, while our method classifies surface points as inaccessible (see \Cref{fig:acc_pinus}). We also validate against mesh-collision ground truth on this tree, which, over an 80,000-point subsample, labels 84.6\,\% of points as accessible, compared with our method's 37.8\,\%, with only 48.0\,\% agreement between the two. This gap is much larger than on the synthetic primitives, where the two ground truths agreed almost exactly, and suggests our method is substantially more conservative on complex branching geometry. Two mechanisms could explain this: the arm-path check's fixed sample distances may behave differently on thin branch structures than on the primitives they were tuned against, and confidence gating may disproportionately suppress labels in dense canopy occlusion.

\subsection{Real world results}
\label{subsec:real_results}

We evaluate our system across four separate Livox Mid-360 scans collected from a field site, containing outdoor tree canopy scenes recorded at walking pace. No ground-truth accessibility labels are available for real LiDAR scans; results are therefore qualitative. \Cref{fig:access_main} shows a representative accessibility field computed over one such scan. The field exhibits the expected spatial structure: outer canopy surfaces and clear branch tips are predominantly accessible, while interior branch geometry, where the tool shaft would have no clear approach corridor, is marked inaccessible.

\textbf{Runtime and memory.}
\Cref{tab:timing} reports per-frame runtime and peak VRAM for our two longest field sequences (3,431 and 3,188 frames) on an RTX 4090 and on a Jetson Orin AGX 32GB embedded platform. At every tested resolution on both devices the pipeline completes within the $100\,\unit{\milli\second}$ LiDAR frame period: the worst-case 95th percentile is $71.4\,\unit{\milli\second}$ at $5\,\unit{\milli\metre}$ on the RTX 4090 and $89.1\,\unit{\milli\second}$ at $10\,\unit{\milli\metre}$ on the Jetson, demonstrating real-time capability on resource-constrained embedded hardware. Peak VRAM is bounded at 1.8--2.1\,GB and 2.8\,GB, respectively, across all resolutions, confirming that the sliding window caps memory use regardless of scene extent.

\begin{table}[h]
  \centering
  \small
  \setlength{\tabcolsep}{4pt}
  \caption{Per-frame pipeline runtime and peak VRAM. RTX~4090 and Jetson Orin AGX 32GB results shown for the two longest field sequences (3.2M and 3.5M accumulated surface points). TSDF integration accounts for under 2\,ms in all cases. Frame budget: 100\,ms.}
  \label{tab:timing}
  \begin{tabular}{llrrr}
    \toprule
    Sequence & Voxel & $\bar{t}$ (ms) & $t^{95}$ (ms) & VRAM (GB) \\
    \midrule
    3.2M pts & 5\,mm  & 42.5 & 71.4 & 2.0 \\
             & 10\,mm & 41.2 & 63.5 & 2.0 \\
             & 20\,mm & 29.7 & 45.7 & 1.8 \\
    \midrule
    3.5M pts & 5\,mm  & 36.0 & 62.2 & 2.1 \\
             & 10\,mm & 38.3 & 66.6 & 2.0 \\
             & 20\,mm & 32.4 & 52.0 & 1.9 \\
    \midrule
    Jetson Orin AGX & 5\,mm  & 41.6 & 81.8 & 2.8 \\
                    & 10\,mm & 59.4 & 89.1 & 2.8 \\
                    & 20\,mm & 49.3 & 73.9 & 2.8 \\
    \bottomrule
  \end{tabular}
\end{table}

\vspace{-0.3cm}
\section{Conclusion}
\label{sec:conclusion}

In this paper, we introduced the Accessibility Field: a per-point labelling of surface accessibility, computed in real time from streaming sparse LiDAR data, that bridges a key gap between scene representation and physical interaction. By combining a scan-centric TSDF integration method with GPU-accelerated tool-geometry kernel queries, we produced this accessibility representation without a prior scene model, at sensor rate for arbitrary tool geometries, on both workstation and NVIDIA Jetson Orin hardware.

Our quantitative evaluation demonstrates that visibility alone is an insufficient proxy for accessibility: on geometry with mixed accessibility, our method achieves F1\,=\,90.8 versus 69.8 for an HPR visibility baseline and 81.5 for the same pipeline with point-cloud collision queries in place of the TSDF, and labels 56.8\,\% of synthetic pine tree branch surfaces as physically inaccessible despite being visible from the sensor. Across the full synthetic suite, three of four scenes achieved F1\,$\geq$\,96.6, with ablative testing confirming that both scan-centric integration and positive TSDF clamping are critical to these results. Real-world qualitative evaluation on Livox Mid-360 field data demonstrates consistent behaviour under the non-repeating scan patterns of sparse outdoor sensors.

\subsection{Limitations and Future Work}
\label{subsec:limit}

The current system limits tool-approach orientation to yaw rotation only, a deliberate simplification suited to our aerial pruning target application, where the UAV cannot pitch forwards while holding position. Extending to full orientation sampling would broaden applicability to arbitrary tool orientations, but at increased query cost. The evaluation presented uses only a single tool geometry per scene, and further controlled study of branch diameter and sensing range is required to isolate the thin-structure benefit of our scan-centric integration. Accessibility field coverage is also constrained by TSDF reconstruction quality: sparse returns on thin structures may produce incomplete surfaces, leading to conservative accessibility estimates in low-density regions. The system assumes a static scene between updates and has not been quantitatively evaluated under scene motion such as wind or branch contact during pruning; the field sequences in this paper were captured under calm conditions, so their consistency does not establish robustness to a moving scene. We treat this as a genuine open limitation, to be addressed via temporal filtering and per-voxel confidence evaluation in future work. Finally, ground-truth accessibility for real LiDAR field scans remains unavailable; while synthetic evaluation provides an initial quantitative validation, our real-scan results are qualitative. Generating pseudo ground-truth from dense multi-view reconstructions is a promising direction for further rigorous real-world evaluation.

\section*{Acknowledgements}
\label{sec:acknowledgements}

The research reported in this article was conducted as part of “Enabling unmanned aerial vehicles (drones) to use tools in complex dynamic environments UOCX2104”, which is funded by the New Zealand Ministry of Business, Innovation and Employment.

\bibliographystyle{named}
\bibliography{references}

\end{document}